\magnification=\magstep1
\baselineskip=16pt
\def\Aplus{${\cal A}\otimes{\cal B}$}

\def\H{\cal H}
\def\D{\cal D}
\def\hh{ \hbox{$ {\rm l} \! {\rm H}$} }
\def\cc{ \hbox{$ {\rm l} \!\!\! {\rm C}$} }

\def\zz{ \hbox{${\rm Z} \!\! {\rm Z}$} }

\vskip2cm
\centerline{\bf REMARKS  ON ALAIN CONNES' APPROACH TO THE STANDARD 
MODEL}
\vskip.2cm
\centerline{\bf IN NON-COMMUTATIVE GEOMETRY}
\vskip1.5cm
\centerline{\bf Daniel Kastler and Thomas Sch\"ucker}
\vskip1cm
\centerline{ Centre de Physique Th\'eorique }

\centerline{ Luminy, case 907 }

\centerline{ F-13288 Marseille, cedex 9 }
\vskip2cm

\centerline{dedicated to the memory of E. M. Polivanov}

\vskip2cm
In the last years Alain Connes produced a remarkable interpretation of the
standard model (electroweak and chromodynamics sectors) within his theory
of non-commutative riemannian spin manifolds [1-4]. All the terms of the
usual bosonic lagrangian are obtained via a non-commutative analogue of
the Yang-Mills algorithm featuring a single curvature attached to a pair
of algebras in Poincar\'e-duality (viz. the electroweak algebra 
${\cal A}=C^\infty\otimes(\cc\oplus\hh)$ and the chromodynamics
algebra ${\cal B}=C^\infty\otimes(\cc\oplus {\bf M}_3(\cc))$ 
building a
``non-commutative space'' \Aplus\  which incorporates the
``inner degrees of freedom''. The differential geometry of 
\Aplus\  is specified (in a way generalizing the 
specification of the differential geometry of a spin$^c$-manifold by its
Dirac operator) by a ``$4^+$-summable $K$-cycle'' (Kasparov 
${\cal A}\otimes{\cal B}$-module) $\H={\H} _{\rm l}\oplus {\H} _{\rm q}$, 
direct sum 
of leptonic
and quark $ \zz/2$-graded Hilbert spaces, also enacted by a generalized 
Dirac operator $\D=\D_{\rm l}\oplus \D_{\rm q}$. The representation 
$\pi=\pi_l\oplus\pi_q$ of 
${\cal A}\otimes{\cal B}$ on the bimodule $\H_{\rm l}\oplus \H_{\rm q}$ 
extends to 
a representation  of the tensor product 
$\Omega_D{\cal A}\otimes\Omega_D{\cal B}$
of their ``non-commutative De Rham complexes'' (sets of 
quantum differential forms). 
The theory produces the four usual gauge bosons plus the Higgs boson 
appearing as a 
fifth gauge
boson connected with discreteness. For details we refer to [5] and [8].

The ``non-commutative Yang-Mills scheme'' consists in ``integrating'' the 
square of the 
quantum 
curvature by means of a trace $\underline{\tau_{\cal D}}$ on the 
endomorphisms of the 
quantum forms
derivate of the trace $\tau_{\cal D}=Tr_{\omega}[{\cal D}^{-4}\pi(\cdot)]$ 
on the 
quantum forms
constructed via Dixmier trace and generalized Dirac operator. In fact, 
as this trace
splits in independent partial traces according to natural splittings of the
Hilbert space $\cal H$, convex combination of those partial traces give rise 
to
a convex family of ``integrations'' (depending upon the adopted splitting). The
 maximal
splitting into irreducible submodules produces the lagrangian of the usual
standard model with its 18 free parameters. A more restricted natural 
choice
corresponds to the splitting $\H={\cal H}_{\rm l}\oplus {\cal H}_{\rm q}$ 
into 
leptonic and quark
Hilbert spaces, with possible traces convex combinations
 $$\alpha_l\tau_{{\cal D}_l}+\alpha_q\tau_{{\cal D}_q},\quad\alpha_l=
{1\over2}(1+x),
\alpha_q={1\over2}(1-x),\quad -1\leq x\leq1,\eqno(1)$$
amongst which the most symmetric choice $x=0$ yields a lagrangian with 4 
parameters less than usual (the free parameters are the entries of the 
fermion
mass matrix plus a single universal coupling constant).

Such constrained versions of the standard model (natural inasmuch as they 
postulate
extents of ``fermionic universality'') are potentially interesting since 
containing more information than the usual standard model
(these models are of the ``unification'' type, with a single Higgs, and without 
cross terms endangering the stability of the proton). Genuine physical 
predictions along such lines would of course require a renormalized 
quantum field theory --- a stage not yet attained in the present purely 
classical
(lagrangian) level, where customary quantization and unsophisticated
renormalization would destroy the constraints and restitute the usual
18 free parameters.

With this in mind, it is, however, perhaps interesting, for a first exploration
of the constrained lagrangians and for speculations, to look at the results at
tree level. The Dixmier trace (1) leads to the following bosonic (gauge and
Higgs) lagrangian [5]:\baselineskip12pt
\footnote{$^1$}{In what follows, we do not bother about the
fact that the traditional lagrangian is lorentzian whilst the NCDG lagrangian
is euclidean, since this point has not bearing on coupling constants.}
\baselineskip16pt
 $${\cal L}_B=-A{{\bf g}^a}_{\mu\nu}{{\bf g}_a}^{\mu\nu}-B{\bf f}_{\mu\nu}
{\bf f}^{\mu\nu}-{1\over4}C{{\bf h}^s}_{\mu\nu}{{\bf h}_s}^{\mu\nu}+
2L{\bf D}_\mu\Phi
_j{\bf D}^\mu\Phi^j+K(\Phi_i\Phi^i-1)^2,\eqno(2)$$
featuring
the curvatures {\bf g,f} and {\bf h} of respective SU(3)-, U(1)-,
and SU(2)-connexions with one-forms {\bf c,a,} and {\bf c}, and the covariant
${\bf D}_\mu\Phi_j$, where 
 $${\bf D}_\mu=\nabla_\mu+i({\bf a}_\mu-{{\bf b}^s}_\mu {\tau_s\over 2}).
\eqno(3)$$  
with the constants $A$ through $K$
\baselineskip12pt\footnote{$^2$}{The 
approximate values for $L$ and $K$ are valid up to 1\%, which is consistent 
with
neglecting all fermion masses against the top mass. }
\baselineskip16pt

 given by
 $$\cases{A={N\over2}(1-x)\cr
\cr
B={N\over3}(10-x)\cr
\cr
C=N(2-x)\cr
\noalign{\vskip.3cm}
L=Tr[\alpha_l\mu_e+3\alpha_q(\mu_u+\mu_d)]\simeq{3\over2}
(1-x)m^2_t\cr
\noalign{\vskip.3cm}
K={3\over2}Tr[\alpha_l\mu_e^2+3\alpha_q(\mu_u^2+\mu_d^2)]+
3\alpha_qTr[\mu_u\mu_d]
\cr\noalign{\vskip.25cm}\quad \quad \quad \quad
-[1/(\alpha_l+6\alpha_q)+1/(2\alpha_l+6\alpha_q)]
N^{-1}L^2
\cr\noalign{\vskip.25cm}\quad \quad \simeq{9\over4}(1-x)m^4_t
-{9\over8}(1-x){{3x^2-8x+5}\over{5x^2-17x+14}}m^4_t\cr}
\eqno(4)$$
\medskip\noindent
where $\mu_e=M_eM_e^*,\mu_u=M_uM_u^*,\mu_d=M_dM_d^*,M_e,M_u,M_d$ 
the respective
mass matrices of the charged leptons, the upper quarks and the lower quarks. 
We
indicated approximate values in terms of the top mass $m_t$ supposed 
dominant.

We recall the expression of the gauge and Higgs part of the lagrangian of the
traditional standard model:
 $$
\eqalign{{\cal L}_{gauge}+{\cal L}_{Higgs}=-{1\over4}{{\cal G}^a}_{\mu\nu}
{{\cal G}_a}
^{\mu\nu}-&{1\over4}B_{\mu\nu}B^{\mu\nu}-
{1\over4}{W^s}_{\mu\nu}{W_s}^{\mu\nu}\cr
&+(D_\mu\phi)^*(D^\mu\phi)+{{\mu^2}\over{
v^2}}(\phi^*\phi-{v^2\over2})^2,\cr}\eqno(5)$$
with the covariant derivative $D_\mu$ given by:
 $$D_\mu=\partial_\mu-i{g_1\over2}B_\mu-ig_2{W^s}_\mu{\tau_s\over2}.
\eqno(6)$$
We find it convenient to use the following basic parameters of the 
standard model --- numerically all known except $\mu$:
 $$\left.\cases{g = \sqrt{g_1^2+g^2_2}\cr
\cos\theta_W=g_2/g\cr
\mu\cr v\cr}\right. .\eqno(7)$$
In terms of the latter, one has $g_1=g\sin\theta_W, g_2=g\cos\theta_W$, 
and the following masses:
 $$\cases{m_H=\sqrt{2}\mu\cr
m_Z={1\over2}vg\cr
m_W=m_Z\cos\theta _W={1\over2}vg_2\cr}\eqno(8)$$
whilst the Higgs potential is given by:
 $$V_{Higgs}={\mu^2\over v^2}
(\phi^*\phi-{v^2\over2})^2.\eqno(9)$$

Identification of the covariant derivatives (3) and (6) is synonimous with
the identifications:
 $$\cases{{\bf c}={g_3}{\cal G}\cr
{\rm in\ components}\ {{\bf c}^a}_\mu= {g_3}{{\cal G}^a}_\mu \quad 
a=1,...,8,\cr}
\eqno(10)$$
\bigskip
 $$\cases{{\bf a}=-{1\over2}g_1B\cr
{\rm in\ components}\ {\bf a}_\mu=g_1B_\mu,\quad \phantom{a=1,...,8,}\cr} 
 \eqno(11)$$
\bigskip
 $$\cases{{\bf b}=g_2 W\cr
{\rm in\ components}\ {{\bf b}^s}_\mu=g_2{ W^s}_\mu,\quad
s=1,2,3.\cr}\eqno(12)$$
implying
 $$\cases{{\bf f}_{\mu\nu}=-{1\over2}g_1B_{\mu\nu}=-{1\over2}g_2
\cos\theta_WB_{\mu\nu}&\cr
\cr
{{\bf h}^s}_{\mu\nu}=g_2{W^s}_{\mu\nu},& s=1,2,3,\cr
\cr
{{\bf g}^a}_{\mu\nu}= {g_3}{{\cal G}^a}_\mu,& a=1,...8\cr}\eqno(13)$$
\medskip \noindent
Assuming that $\phi$ and $\Phi$ differ by a constant (insensitive to 
multiplication
of $\phi$ resp. $\Phi$, by constants), the latter follows from comparison of
the fourth terms of (2) and (5), yielding
 $${v\over{\sqrt2}}\Phi=\phi.\eqno(14)$$
Inserting (10) through (14) into (2) yields:
 $$\eqalign{
C^{-1}g_2^{-2}{\cal L}_B=-{A\over C}{{ {g_3}^2}\over{g^2_2}}{{\cal G}^a}_
{\mu\nu}
{{\cal G}_a}^{\mu\nu}-&{B\over{4C}}\tan^2\theta_WB_{\mu\nu}B^{\mu\nu}
-{1\over4}{W^s}_{\mu\nu}{W_s}^{\mu\nu}\cr
&+{{4L}\over{Cv^2g^2_2}}\left\lbrace
(D_\mu\phi)^*(D^\mu\phi)+{K\over{v^2L}}
\left(\phi^*\phi-{v^2\over2}\right)^2\right\rbrace,
\cr}\eqno(15)$$
which we identify with (5). Identification of the first term yields:
 $${g_3}={1\over2}(C/A)^{1/2}g_2.\eqno(16)$$
Identification of the second terms fixes the Weinberg angle:
 $$\tan^{-2}\theta_W={B\over C}\quad{\rm whence}\quad 
\sin^2\theta_W={C\over{B+C}}
\eqno(17)$$
Identification of the middle terms fixes the ratio of the mass $m_W$ of the
$W$-boson to the mass $m_t$ of the top quark:
 $$v^2g^2_2=4m_W^2=4{L\over C}\eqno(18)$$
whence
 $$m_W=(L/C)^{1/2}.\eqno(18a)$$

Identification of the ratio of the last terms fixes the ratio of the mass
$m_H$ of the Higgs boson to the mass $m_t$ of the top quark:
 $$\mu^2=K/L,\eqno(19)$$
whence
 $$m_H=(2K/L)^{1/2}.\eqno(19a)$$

Plugging (3) into those relations gives

 $${g_3}={1\over2}[{{4-2x}\over{1-x}}]^{1/2}g_2\quad (=g_2\quad
 {\rm for}\quad x=0),\eqno(20)$$
 $$\sin^2\theta_W={{3(1-x/2)}\over{8-2x}}\quad (={3\over8}\quad {\rm for}
\quad x=0),\eqno(21)$$
and, {\it neglecting all fermion masses against the top mass} $m_t$:
 $$m_W=\left({{3(1-x)}\over{4N(1-x/2)}}\right)^{1/2}m_t
 \quad (={1\over2}m_t\quad{\rm for}\quad x=0\quad {\rm and}\quad N=3)
\eqno(22)$$
and
$$m_H=\left(3-{3\over2}{{3x^2-8x+5}\over{5x^2-17x+14}}\right)^{1/2}m_t.
\eqno(23)$$
Relations (20) and (21) have a ``grand unification'' flavour. We now show
that, given the relationships (10, (11), (12) between fields and connexion
forms, those relations proceed directly from the choice (1) of Dixmier trace
together with the gauge content of the leptonic and quark $K$-cycle
${\cal H}_l$ and ${\cal H}_q$. Concerning the latter  we  recall the 
definitions of the leptonic and quark Hilbert spaces (cf. [4],[5]): we have 
${\cal H}_l=L^2({\bf S_M})\otimes H_l$ and 
${\cal H}_q=L^2(({\bf S_M})\otimes H_q$, where ${\bf S_M}$ is the spin 
bundle,
and:
 $$H_l={H_l}_R\oplus {H_l}_L\eqno(24)$$
with
\baselineskip12pt
\footnote{$^3$}{The factor \  ${\cc}^N$ corresponds to the $N$ fermion 
generation
with identical gauge behaviour, and can be omitted in the calculation to 
come.}
\baselineskip16pt 
$$\cases{H_{l_R}={\cc}^1_R\otimes{\cc}^N,& ${\cc}^1_R$\ spanned\ by
\ $e_R$,\cr
\cr
H_{l_L}={\cc}^2_L\otimes{\cc}^N,& ${\cc}^1_R$\ spanned\ by
\ $\nu_L,e_L$\cr},\eqno(25)$$
respectively:
 $$H_q=H_{q_R}\oplus H_{q_L}\eqno(26)$$
with
  $$\cases{H_{q_R}={\cc}^2_R\otimes{\cc}^N,& ${\cc}^2_R$\ spanned\ by
\ $u_R,d_R$,\cr
\cr
H_{q_L}={\cc}^2_L\otimes{\cc}^N,& ${\cc}^2_L$\ spanned\ by
\ $u_L,d_L$,\cr}\eqno(27)$$
The leptonic inner space ${\cc}^1_R\oplus{\cc}^2_l$ is acted upon
by matrix representatives $T^3_l,C^3_l, Q_l$ of $T^3$, $C^3$,
and of the electric charge Q
\baselineskip12pt
\footnote{$^4$}
{We choose to 
work with the electric charge $Q$ for the definition of which there is
consensus, rather than with the hypercharge $Y$ which is 
differently defined by different authors. We shall need the infinitesimal 
generator $R$ of the $U(1)$ group to write formula (34) below, but we shall
 not 
really need it.}
\baselineskip16pt
, where $iT^3\in  su(2)$, $iC^3\in su(3)$:

 $$T^3_l=\pmatrix{0&0&0\cr  0&{1\over2}&0\cr  0&0&-{1\over2}\cr}
\matrix{e_R\cr \nu_L\cr e_L\cr}\quad s.t.\quad Tr(T^{3*}_lT^3_l)=
{1\over2},\eqno(28)$$

\bigskip
 $$C^3_l=0\eqno(29)$$
\bigskip
 $$Q_l=\pmatrix{-1&0&0\cr  0&0&0\cr  0&0&-1\cr} \matrix{e_R\cr  \nu_L 
\cr e_L\cr}
\quad s.t. \quad Tr(Q_l^*Q_l)=2.\eqno(30)$$
\bigskip
Likewise the quark inner space ${\cc}^2_R\oplus {\cc}^2_L$ is acted upon
by representatives $T^3_q,C^3_q,Q_q$ of $T^3,C^3$, and $Q$, with
matrices:
 $$T^3_q=\pmatrix{0&0&0&0\cr
0&0&0&0\cr  0&0&{1\over2}&0\cr  0&0&0&-{1\over2}\cr}\matrix{u_R\cr 
d_R\cr u_L\cr d_L\cr}
\otimes\pmatrix{1&0&0\cr 0&1&0\cr 0&0&1\cr}\quad s.t. 
\quad Tr(T^{3*}_qT^3_q)={3\over2},
\eqno(31)$$
\bigskip
 $$C^3_q=\pmatrix{1&0&0&0\cr 0&1&0&0\cr 0&0&1&0\cr 0&0&0&1\cr}
\matrix{u_R\cr d_R\cr u_L\cr d_L\cr}\ \otimes\ 
\pmatrix{{1\over2}&0&0\cr 
0&-{1\over2}&0\cr
0&0&0\cr}\quad s.t. \quad Tr(C^{3*}_qC^3_q)=2\eqno(32)$$
\bigskip
 $$Q_q=\pmatrix{{2\over3}&0&0&0\cr 0&-{1\over3}&0&0\cr 0&0
&{2\over3}&0\cr
0&0&0&-{1\over3}\cr} 
\matrix{u_R\cr d_R\cr u_L\cr d_L\cr} \ \otimes\ \pmatrix{1&0&0
\cr 0&1&0\cr 
0&0&1\cr}
\quad s.t. \quad Tr(T^{3*}_qT^3_q)={{10}\over3}.\eqno(33)$$
\bigskip

Invariant bilinear forms on the Lie algebras $su(3), su(2)$, and $u(1)$ are
unique up to scale. The most general invariant bilinear form on the Lie 
algebra
$su(3)\oplus su(2)\oplus u(1)$ of the gauge group is therefore an arbitrary 
linear
combination of the type:
 $$\eqalign{
\langle A,A'\rangle={1\over{{g_3}^2}}Tr(C^*C')+&{1\over{g^2_2}}Tr(T^*T')+
{1\over{g_1^2}}
{1\over2}\bar RR',\cr
&A=(C,T,R),A'=(C',T',R')\in su(3)\oplus su(2)\oplus u(1)\cr
\cr}\eqno(34)$$
Identifying this with the bilinear form on the Lie algebra of the gauge
group stemming from the Dixmier trace (1):
 $$\langle A,A'\rangle=\alpha_lTr(A_l^*A'_l)+\alpha_qTr(A_q^*A'_q),\quad 
A,A'\in su(3)\oplus su(2)\oplus u(1),\eqno(35)$$
then yields the relation
 $$\left({{{g_3}}\over{g_2}}\right)^2={{\alpha_lTr(T^{3*}_lT^3_l)
+\alpha_qTr(T^{3*}_qT^3_q)}\over
{\alpha_lTr(C^{3*}_lC^3_l)+\alpha_qTr(C^{3*}_qC^3_q)}}={{{1\over2}
\alpha_l
+{3\over2}\alpha_q}\over{2\alpha_q}}={{2-x}\over{2(1-x)}},\eqno(36)$$
identical with (20), and the relation [9]
 $$\sin^2\theta_W={{\alpha_lTr(T^{3*}_lT^3_l)+\alpha_qTr(T^{3*}_qT^3_q)}
\over
{\alpha_lTr(C_l^*C_l)+\alpha_qTr(Q_q^*Q_q)}}={{{1\over2}\alpha_l
+{3\over2}\alpha_q}\over{2\alpha_l+{{10}\over 3}\alpha_q}}={3\over4}{{2-x}
\over{4-x}},\eqno(37)$$
identical with (21). For $x=0$, this calculation is formally identical to the 
one performed in $SU(5)$-grand unification, the reason being that
fermion content and ponderation are the same.

Our second concern is a short discussion of the behaviour or relations (20) 
through
(23) as $x$ ranges from -1 to +1. This is illustrated by the table:
\vskip.5cm
 $$\vbox{\halign{\hfill#\hfill&\quad\hfill#\hfill&\quad\hfill#\hfill&\quad
\hfill#\hfill&\quad\hfill#\hfill&\quad\hfill#\cr
$x$&-1&0&${1\over2}$&0.99&1\cr
\noalign{\medskip}
$\left({{g_3}/ g_2}\right)^2$&${3\over4}$&1&${3\over2}$&50.5&$\infty$\cr
\noalign{\medskip}
$\sin^2\theta_W$&${9\over{20}}$&${3\over8}$&${9\over{28}}$&0.252&${1
\over4}$\cr
\noalign{\medskip}
${m_t/ m_W}$&$\sqrt 3$&2&$\sqrt 6$&14.2&$\infty$\cr
\noalign{\medskip}
${m_H/m_W}$&$2.65$&$3.14$&$3.96$&$24.5$&$\infty$\cr}}$$
\bigskip
\noindent We note that the ratio $m_H/m_t$ shows little variationfrom 
1.53 to $\sqrt 3$. 
The table suggests the following remarks:
\vskip.3cm
\noindent
(i): all tabulated functions are monotonic in $x$.

\noindent
(ii): the value $x=0$ seems to correspond to a situation of the ``unification''
type.

\noindent
(iii): for the limit value $x=1$, i.e. $\alpha_q=0$, the Weinberg angle is near 
its experimental
value, whilst strong interactions prevail. Indication of lepton dominance
at experimental energies? Connected with confinement?

\bigskip
The first-named author had the privilege, in the seventies, to be amicably
introduced by E. M. Polivanov to some of the architectonic treasures of 
Moscow. D. K.'s lasting impressions about the beauty and spirituality of
the old Russian capital remain indelibly tied up with the remembrance of 
the
kindness and moral elevation of the eminent physicist for whom we are 
mourning.
\bigskip
\centerline{REFERENCES}
\medskip
\item{[1]} A. Connes, The action function in non-commutative geometry, 
Comm. Math. Phys.
{\bf 117}, 673 (1988).
\item{[2]} A. Connes, Essay on Physics and Non-commutative Geometry, 
{\it The Interface
of Mathematics and Particle Physics}, Clarendon Press, Oxford (1990).
\item{[3]} A. Connes and J. Lott, Particle models and non-commutative 
geometry,
Nucl. Phys. B {\bf 18B} (Proc. Suppl.) (1990).
\item{
[5]} A. Connes, The metric aspect of non-commutative geometry. College de
France Preprint (October 1991).
\item{
[8]} D. Kastler, A detailed account of Alain Connes' version of the standard
model in non-commutative geometry, I (1991) and II (1992).
\item{
[9]} T. Sch\"ucker, $Q=I_3+{1 \over 2}Y$, why? Proc. XI Intern. Coll. on 
Group-Theoret.
Methods in Physics, Istanbul, eds. M. Serdaroglu, E. Inonu, Springer, 
Heidelberg (1982).

\end